\documentclass[journal]{IEEEtran}
\usepackage[a4paper, margin=1in]{geometry}
\usepackage{amsmath, amssymb}
\usepackage{graphicx}
\usepackage{hyperref}
\usepackage{authblk}
\usepackage{enumitem}
\usepackage{lipsum}
\usepackage{physics}
\usepackage{booktabs}
\usepackage{cite}
\usepackage{lettrine}

\usepackage{xcolor}

\title{Polarization Dynamics in VCSELs Under Sinusoidal Signal Modulation around the Polarization Switching point}

\author{Salah Guessoum, Tushar Malica, Athanasios Kyriazis, J\"urgen Van Erps, Geert Van Steenberge and Martin Virte. \thanks{This work was supported in part by the Research Foundation Flanders (FWO, Grant number G020621N ), in part by the the European Research Council (ERC, Starting Grant COLOR’UP 948129, MV) and in part by the METHUSALEM program of the Flemish government. \textit{(Corresponding author: Salah Guessoum, salah,eddine.guessoum@vub.be)}} \thanks{S. Guessoum and A. Kyriazis are with Brussels Photonics (B-PHOT), Vrije Universiteit Brussel, Pleinlaan 2, 1050 Brussels, Belgium, also with Flanders Make at VUB - BP$\&$M, 1050 Brussels, Belgium, and also with the Center for Microsystems Technology (CMST), Ghent University and imec, 9052 Ghent, Belgium.} \thanks{T. Malica and Martin Virte are with Brussels Photonics (B-PHOT), Vrije Universiteit Brussel, Pleinlaan 2, 1050 Brussels, Belgium.} \thanks{J. Van Erps is with with Brussels Photonics (B-PHOT), Vrije Universiteit Brussel, Pleinlaan 2, 1050 Brussels, Belgium, also with Flanders Make at VUB - BP$\&$M, 1050 Brussels, Belgium} \thanks{G. Van Steenberge is with the Center for Microsystems Technology (CMST), Ghent University and imec, 9052 Ghent, Belgium}\thanks{All data and materials necessary to reproduce the results are available in Zenodo repository at https://doi.org/10.5281/zenodo.18456280.}}
 
\begin{document}

\maketitle

\begin{abstract}
%SEG
Vertical-Cavity Surface-Emitting Lasers (VCSELs) combine compact geometry, low threshold current, and ease of integration, making them central to modern photonic systems. However, their polarization behavior remains a critical factor affecting performance, as the emission state can switch between orthogonally polarized modes around the so-called polarization switching point. This regime exhibits high sensitivity, where small perturbations induce abrupt polarization changes and nonlinear responses. In this work, the polarization dynamics of VCSELs under sinusoidal current modulation around the switching point are numerically investigated using the Spin-Flip Model. The study examines the influence of modulation frequency, amplitude, and bias current, revealing distinct dynamical regimes including polarization locking, periodic and irregular switching. The observed transitions between regimes elucidate the interplay between modulation and polarization stability, providing insight into the control of VCSEL dynamics for high-speed optical communication and sensing applications.
\end{abstract}

\section{Introduction}
\lettrine[lines=2,lhang=0,nindent=0em]{V}{ertical}-cavity surface-emitting lasers (VCSELs) are key components in optical interconnects, sensing, and quantum systems thanks to their compact size, low threshold currents, circular beam profiles, and ease of integration \cite{Iga_Review_2008, Michalzik_VCSELs_2013,Iga_VCSELHistory,Larsson_AdvancesVCSEL2011}. Besides their role as light sources, VCSELs offer an interesting degree of freedom in the form of their polarization state, arising from the near-degeneracy of two orthogonal linearly polarized (LP) cavity modes \cite{Van_Exter98:Polarization_Fluctuation_theory,Regalado1997:VCSEL_Polarization_Properties}. The competition between these modes leads to rich nonlinear polarization dynamics, especially near the polarization switching (PS) threshold, where the system becomes highly sensitive to perturbations and external forcing \cite{VCSEL_fundamentals_book_Tatum2021}.\\
Many VCSEL applications benefit from the versatility provided by controlling the polarization state. Examples include short-reach optical communication links \cite{Priyadarshi:2012_Condit_Chaotic_Comm_Ext_cav_laser, Gatto:16_High_Capacitance_VCSEL_Short_reach_comm}, optical sensing and spectroscopy \cite{Liu:19_VCSEL_applications_comm_sens}. In these applications, uncontrolled polarization fluctuations can produce intensity noise, pattern distortions, reduced signal accuracy, and degraded measurement contrast or stability.\\
Despite extensive studies, the mechanisms driving polarization switching are not fully understood as the interplay between deterministic nonlinear effects and stochastic fluctuations remains an open question \cite{Danckaert_Stochastic_dynamics,M.Virte2013:Bifurcation_Pol_Dyn_Chaos}. Experiments have shown that PS often coincides with spontaneous mode hopping \cite{Willemsen:1999_PSModeHopping} and noise-driven transitions \cite{M.Jadan:VCSELBistab2019}, confirming the strong role of stochasticity in this regime. At the same time, chaotic dynamics have also been identified as a part of the polarization switching process \cite{M.Virte2013:Bifurcation_Pol_Dyn_Chaos, virte_deterministic_2013}. These findings suggest that VCSELs behave as nonlinear bistable systems influenced by both intrinsic dynamics and external noise.
On the other hand, several external mechanisms can trigger polarization switching by disturbing the balance between the two LP modes. Reported approaches include modulation of the injection current \cite{Paul07:Dynamical_hysterisis,Quirce:24_PolFluctuation_TurnOn,Masoller06:Mod_Pol_Switch_Sweeps,Temgoua_2021:AM-FM_Modulation,Xu_2023_Pol_SpinVCSEL}, optical pumping \cite{M.LindemannUltrafastSpin_VCSEL:2019,Lindemann2023:Pol_Dyn_VCSEL_Grating,Tselios_Mod_QD_VCSEL:23,Denis-leCoarer_Mode_hopping_optical_parallel_inj:2018}, and delayed optical feedback \cite{Chen_OIL_PolarizationSwitching,NAZHAN:2015_PS_Optical_feedback_modulated}. Each method perturbs the internal competition, either by modifying carrier dynamics or by introducing phase and amplitude asymmetries, enabling controlled polarization transitions. \\
While polarization switching has been investigated under a variety of perturbations, the role of deterministic periodic forcing remains comparatively less explored. In particular, sinusoidal current modulation represents a technologically relevant mechanism, since it directly connects to the way VCSELs are driven in practical high-speed communication systems. By tuning modulation amplitude and frequency, one can potentially access different dynamic regimes, ranging from smooth polarization oscillations to abrupt switching and even chaotic responses. A systematic study of this behavior is thus essential for advancing the fundamental understanding of VCSEL nonlinear dynamics as well as for identifying design strategies for polarization-stable devices in emerging photonic platforms, such as integrated photonics \cite{Pan2024_Review_VCSEL_PICs}, and VCSEL-based holographic systems \cite{Yang_Hui:2022_VCSEL_Tri_dimension_holography}.\\
In this work, we investigate the response of VCSELs to sinusoidal current modulation close to polarization switching conditions. We focus on how modulation frequency and amplitude affect the polarization dynamics. Our analysis is based on the Spin-Flip Model (SFM), which accounts for spin-carrier interactions, birefringence and anisotropies, and provides insight into the polarization behavior of VCSELs near the switching threshold.
%bottom is commented following Jurgen comment saying this has been mentioned prior to this 
% These results are relevant for optimizing VCSEL performance in optical communication and sensing applications. 

\section{Simulation model and parameters}
The polarization dynamics of a VCSEL can be described by the Spin-Flip Model (SFM) \cite{SanMiguel_Feng_Moloney_95}. The model tracks the evolution of the right- and left-handed circularly polarized field amplitudes \(R_{\pm}\), their relative phase \(\phi = \Psi_+ - \Psi_-\), the total carrier density \(N\), and the carrier spin imbalance \(S\). Expressed in amplitude and phase form, the equations are:

\begin{align}
\frac{dR_+}{dt} &= \kappa (N + S - 1) R_+ - (\gamma_a \cos \phi + \gamma_p \sin \phi) R_-, \\
\frac{dR_-}{dt} &= \kappa (N - S - 1) R_- - (\gamma_a \cos \phi - \gamma_p \sin \phi) R_+,  \\
\frac{d\phi}{dt} &= 2\kappa \alpha S - \left( \frac{R_-}{R_+} - \frac{R_+}{R_-} \right) \gamma_p \cos \phi \nonumber \\
&\quad + \left( \frac{R_-}{R_+} + \frac{R_+}{R_-} \right) \gamma_a \sin \phi,  \\
\frac{dN}{dt} &= -\gamma \left[ N - \mu(t) + (N + S) R_+^2 + (N - S) R_-^2 \right],  \\
\frac{dS}{dt} &= -\gamma_s S - \gamma \left[ (N + S) R_+^2 - (N - S) R_-^2 \right].
\end{align}

The current modulation is introduced in the injection current term: \(\mu(t) = \mu_0 (1 + A \sin(2\pi f t)),\) where \(\mu_0\) is the bias current, \(f\) is the modulation frequency and \(A\) is the modulation amplitude, defined here as the decimal ratio of the selected current \(\mu_0\). Note that the injection current is normalized to the threshold \(\mu_{\text{th}}\) which makes \(\mu(t)\) dimensionless. \(\kappa\) and \(\alpha\) are defined in table~\ref{tab:params}. The phase and amplitude anisotropies have been set to $\gamma_a = -0.5 \, ns^{-1}$ and $\gamma_p=10 \, ns^{-1}$ to obtain polarization switching at low injection current  \(1 \leq \mu_{\text{ps}} \leq 2.5\),  matching realistic operating conditions \cite{Masoller06:Mod_Pol_Switch_Sweeps,M.Virte2013:Bifurcation_Pol_Dyn_Chaos}. 

\begin{figure}[h!]
\centering
\includegraphics[width=0.9\linewidth]{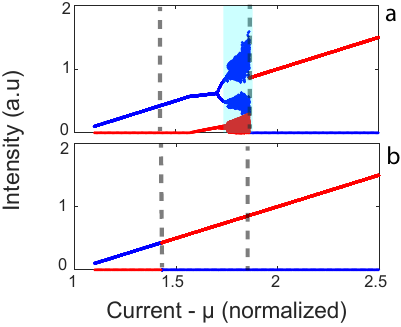}
\caption{Polarization-resolved output power under DC bias sweeps. (a) Increasing and (b) decreasing bias current sweeps show a hysteresis loop that marks the polarization switching region between X-LP and Y-LP modes. The highlighted region in panel (a) indicates the emergence of complex dynamical behavior associated with a change in the underlying attractor during the switching mechanism.}
\label{fig:Bifurcation_diag}
\end{figure}

The polarization switching region is identified using steady-state stability diagrams, where PS occurs at the boundary between the X- and Y-linearly polarized modes. Figure~\ref{fig:Bifurcation_diag} shows the simulated polarization-resolved output power under DC bias sweeps, confirming bistable behavior with hysteresis between \(\mu = 1.48\) and \(\mu = 1.78\). Within this interval, the increasing-current sweeps (Fig.~\ref{fig:Bifurcation_diag}.a) reveal a destabilization of the relaxation-oscillation dynamics, leading to complex non-periodic dynamics, consistent with a secondary destabilization of the relaxation-oscillation limit cycle. This is indicative of attractor competition between polarization modes, whereas the reverse sweep (Fig.~\ref{fig:Bifurcation_diag}.b) exhibits a sudden switching without any instabilities.\\
The rest of the laser parameters are listed in Table~\ref{tab:params} and are kept fixed in our study. 
To analyze the VCSEL dynamics, we use time traces of the linearly polarized components of the VCSEL \(E_x\) and \(E_y\), given respectively by: $E_x = \tfrac{1}{\sqrt{2}}\left(E_{+} + E_{-}\right)$ and 
$E_y = \tfrac{i}{\sqrt{2}}\left(E_{+}- E_{-} \right)$.

\begin{table}[h!]
\centering
\caption{Simulation parameters used in the Spin-Flip Model}
\label{tab:params}
\begin{tabular}{llc}
\toprule
\textbf{Parameter} & \textbf{Description} & \textbf{Value} \\
\midrule
\(\alpha\) & Linewidth enhancement factor & 3 \\
\(\gamma\) & Carrier decay rate & \(1~\mathrm{ns}^{-1}\) \\
\(\gamma_s\) & Spin-flip relaxation rate & \(100~\mathrm{ns}^{-1}\) \\
\(\kappa\) & Field decay rate & \(600~\mathrm{ns}^{-1}\) \\
\(\gamma_a\) & Amplitude anisotropy & \(-0.5~\mathrm{ns}^{-1}\) \\
\(\gamma_p\) & Birefringence &  \(10~\mathrm{ns}^{-1}\) \\
\bottomrule
\end{tabular}
\end{table}

\begin{figure}
\centering
\includegraphics[width=\linewidth]{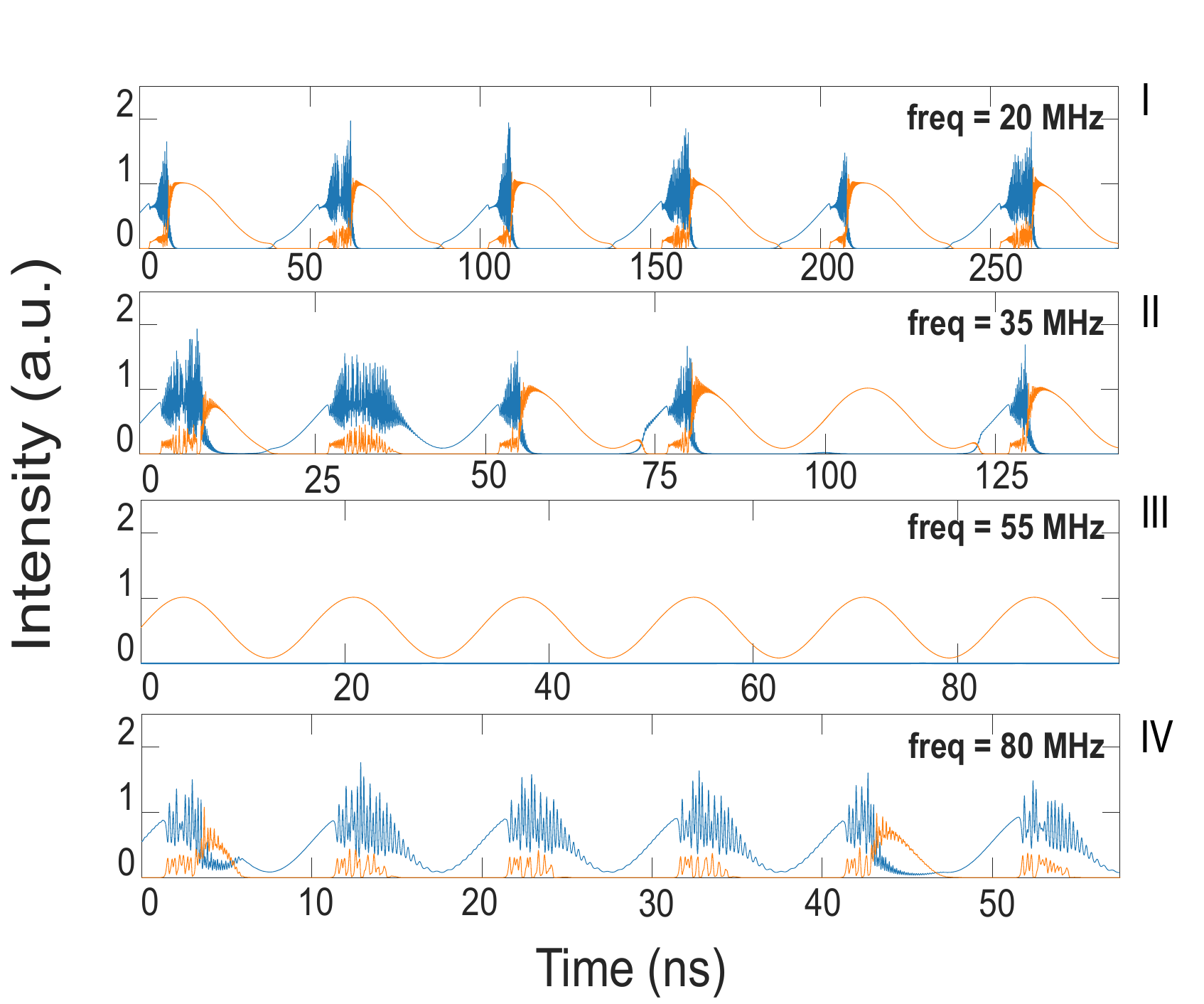}
\caption{Temporal evolution of the polarization state of the X-LP (blue) and Y-LP (orange) under sinusoidal modulation, illustrating the four dynamical regimes: (I) deterministic switching, (II) partial switching, (III) polarization locking, and (IV) stochastic behaviour. Each subplot corresponds to a different modulation frequency, with constant modulation amplitude and bias current. The horizontal timescale is adjusted to display a similar number of periods.}
\label{fig:Time-Series}
\end{figure}

\section{Dynamical Regimes under current modulation}
Applying a sinusoidal current modulation around the polarization switching point, we expect to trigger regular switches at low frequencies and only observe small oscillations above a certain frequency threshold due to the limited modulation bandwidth of the laser, typically on the order of a few gigahertz for standard VCSELs. In practice, we observe different complex dynamic regimes, most of which show little consistency compared to the periodic stimulus. With current $\mu = 1.55$ and an oscillation amplitude of $0.3$, we report distinct regimes, displayed in Fig. \ref{fig:Time-Series} showing time-series for the X and Y linear polarizations:\\

\textbf{Regime (I)} — Systematic polarization switching synchronized with modulation signal. At low modulation frequencies (below 30 MHz), the laser exhibits clean, repeatable polarization switches synchronized with the drive signal. The trajectory follows the hysteresis loop, alternating between the X- and Y-polarized states. We observe short dynamical transitions, consistent with the dynamical behavior visible in the bifurcation diagram. Yet, we already see that the behavior of the laser is not strictly periodic but this can easily be attributed to the chaotic behavior accompanying the polarization switching shown in Fig.~\ref{fig:Bifurcation_diag} \cite{M.Virte2013:Bifurcation_Pol_Dyn_Chaos}. 

\textbf{Regime (II)} — Non-systematic and irregular switching with Y-LP dominance: At moderate frequencies (30–50 MHz), switching becomes non-systematic and irregular. The system develops a preference for Y-LP, with asymmetric transitions and missed switches. Although the X-LP mode sometimes remains for a full cycle, we mainly observe the emergence of a clean sine period on Y-LP with suppressed X-LP. Such an occurrence becomes gradually more frequent as the frequency is increased until it reaches regime (III). 

\textbf{Regime (III)} — Single polarization modulation, no switching: At intermediate frequencies (50–70 MHz), switching is fully suppressed and the laser becomes locked on the Y-LP linear polarization. Although the modulation amplitude spans the same bistable range, the system doesn't follow the hysteresis cycle. It looks as if the modulation was too fast for the laser to break free from the basin of attraction of the Y-LP steady-state.

\textbf{Regime (IV)} — rare switching with dynamics and dominant X-LP: above 70 MHz, the laser starts exhibiting irregular switching events, with an increasing fraction of periods dominated by the X-LP mode rather than the Y-LP mode. Above 200 MHz, X-LP is almost systematically dominant and Y-LP only sporadically appears. Dynamics are clearly visible on both modes and for all periods. 

\textbf{Regime (V)} — not shown in Fig. \ref{fig:Time-Series}. At high modulation frequencies in the GHz range, we observe emission from Y-LP only, similar to Regime (III). However, the oscillation amplitude gradually decreases as the frequency exceeds the modulation bandwidth of the laser. This regime is consistent with standard laser behavior and is therefore left out of the rest of our investigations in this work.\\
\begin{figure}
\centering
\includegraphics[width=\linewidth]{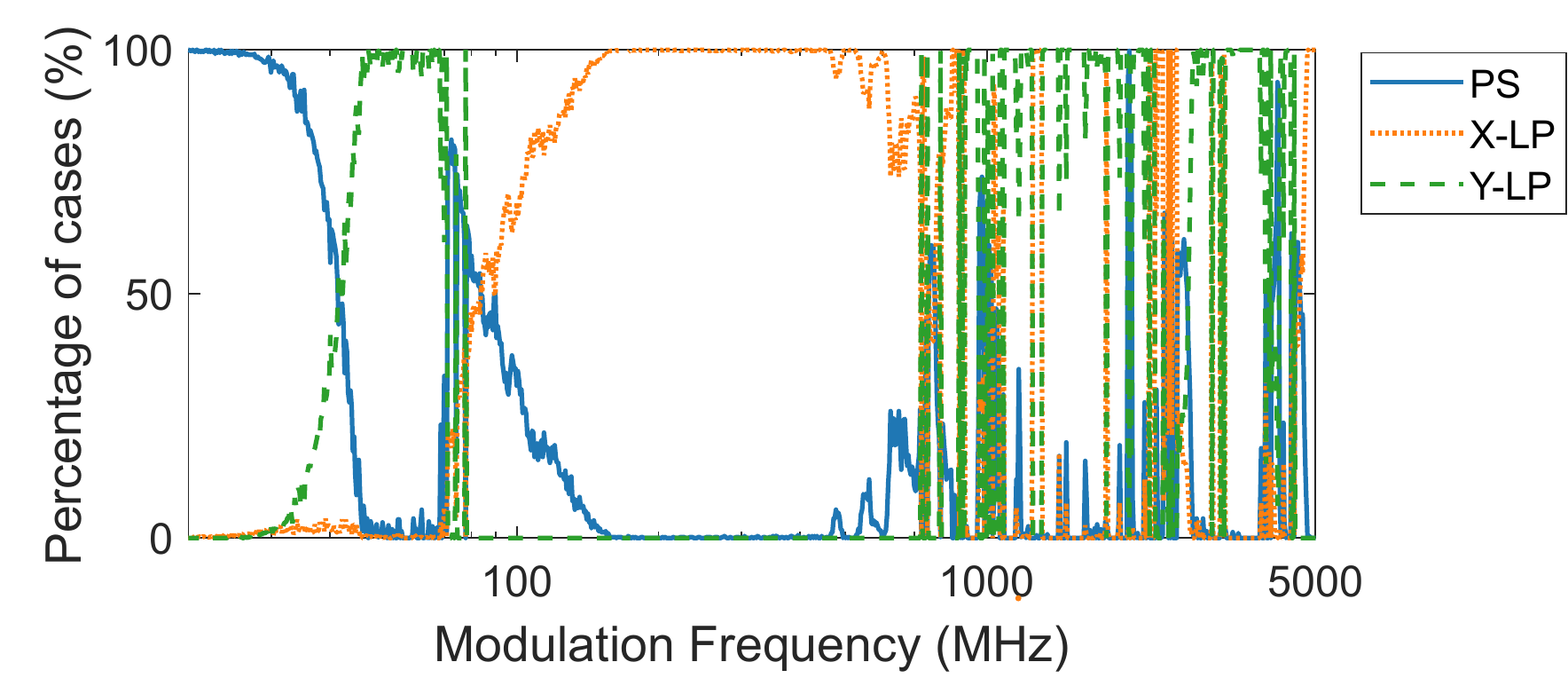}
\caption{Probability of polarization switching (PS), X-LP dominance and Y-LP dominance as a function of modulation frequency at \(\mu = 1.55\). Dominance is defined as exceeding 70\% of the total output power.}
\label{fig:Percentage_155}
\end{figure}
For a more systematic analysis, we classify the dynamics of each modulation cycle as either (i) polarization switching (both X-LP and Y-LP are dominant during a certain part of the cycle), (ii) X-LP dominance (no switching and more than 70\% of total power contained in X-LP), or (iii) Y-LP dominance (no switching and more than 70\% of the total power contained in Y-LP); The 70\% threshold is defined arbitrarily. For each modulation frequency, we simulate a thousand modulation cycles and assign a classification to every individual cycle according to the above criteria. The evolution of the relative occurrence (percentage of cases) of each class is shown in Fig.~\ref{fig:Percentage_155}. From this distribution, we clearly observe that regimes (II) and (III) represent an initial transition in which Y-LP progressively becomes dominant. Around 70 MHz, the laser appears to cross a threshold, marked by the re-emergence of the X-LP mode, which becomes largely dominant again above 100 MHz. Closer to 1 GHz, a second transition occurs, characterized by sporadic Y-LP dominance but without a persistent trend.
\begin{figure*}[tb]
\centering
\includegraphics[width=\textwidth,height=10cm]{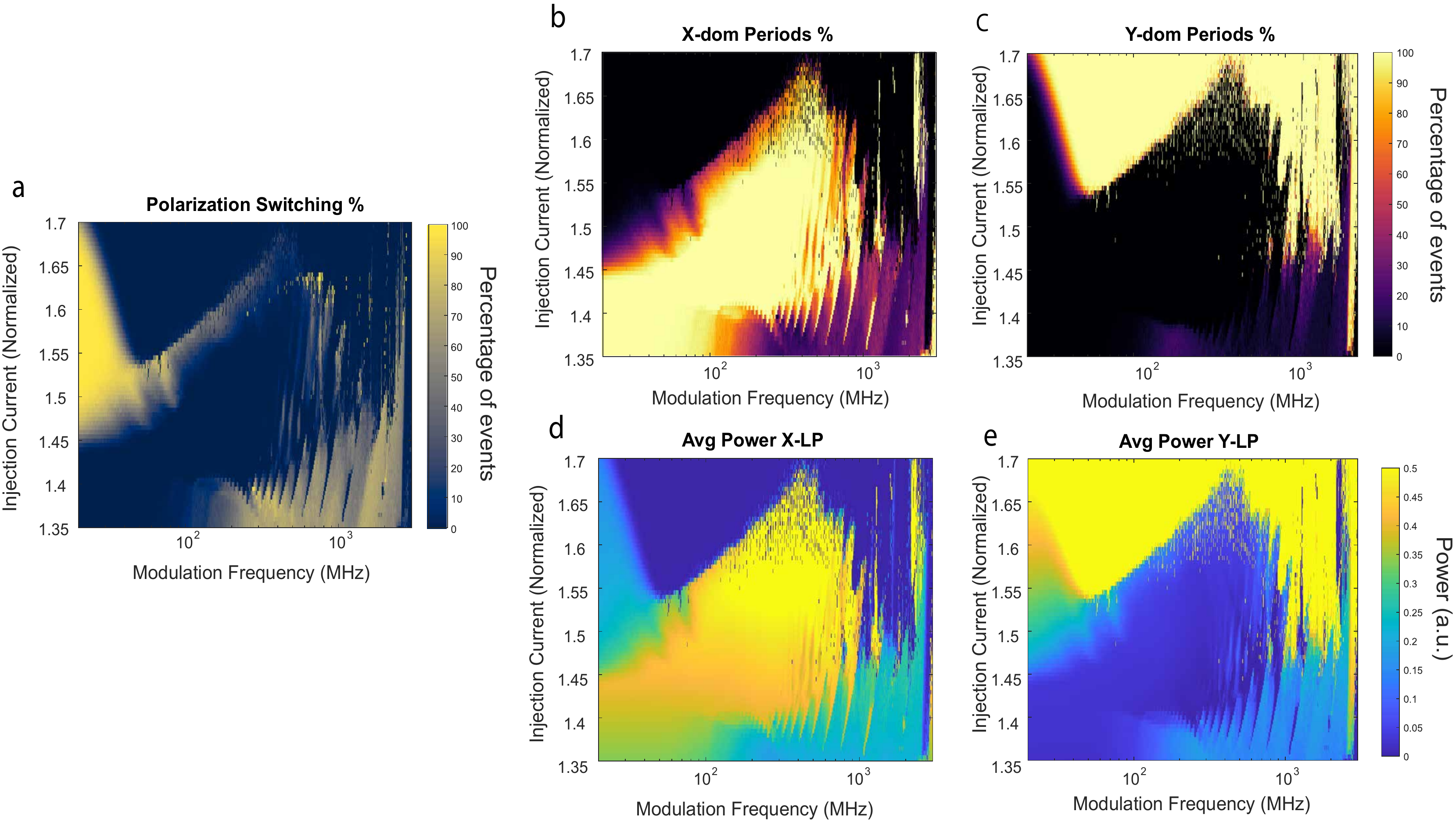}
\caption{Colormaps summarizing the VCSEL response as a function of modulation frequency and injection current at a modulation amplitude of 0.3. Colormaps a, b, and c show respectively the polarization switching percentage, the fraction of X-LP dominance, and the fraction of Y-LP dominance. Colormaps d and e show, respectively, the average output powers of the X-LP and Y-LP components.}
\label{fig:Probability_colormap}
\end{figure*}
At first, regimes (II) and (III) look like the laser is simply reaching a certain limit of its modulation bandwidth: the modulation frequency is too high and thus the laser only sees a smaller amplitude, which keeps it stuck on the Y-LP mode. However, the emergence of a fourth regime, especially combined with the low value of this threshold frequency, below 100 MHz in the example shown, while a modulation bandwidth is typically in the GHz range, rather calls for a different interpretation related to the nonlinear dynamics of the laser and the polarization mode competition.

\section{Influence of injection current offset and modulation amplitude}
\begin{figure}
  \centering
  \includegraphics[width=\linewidth]{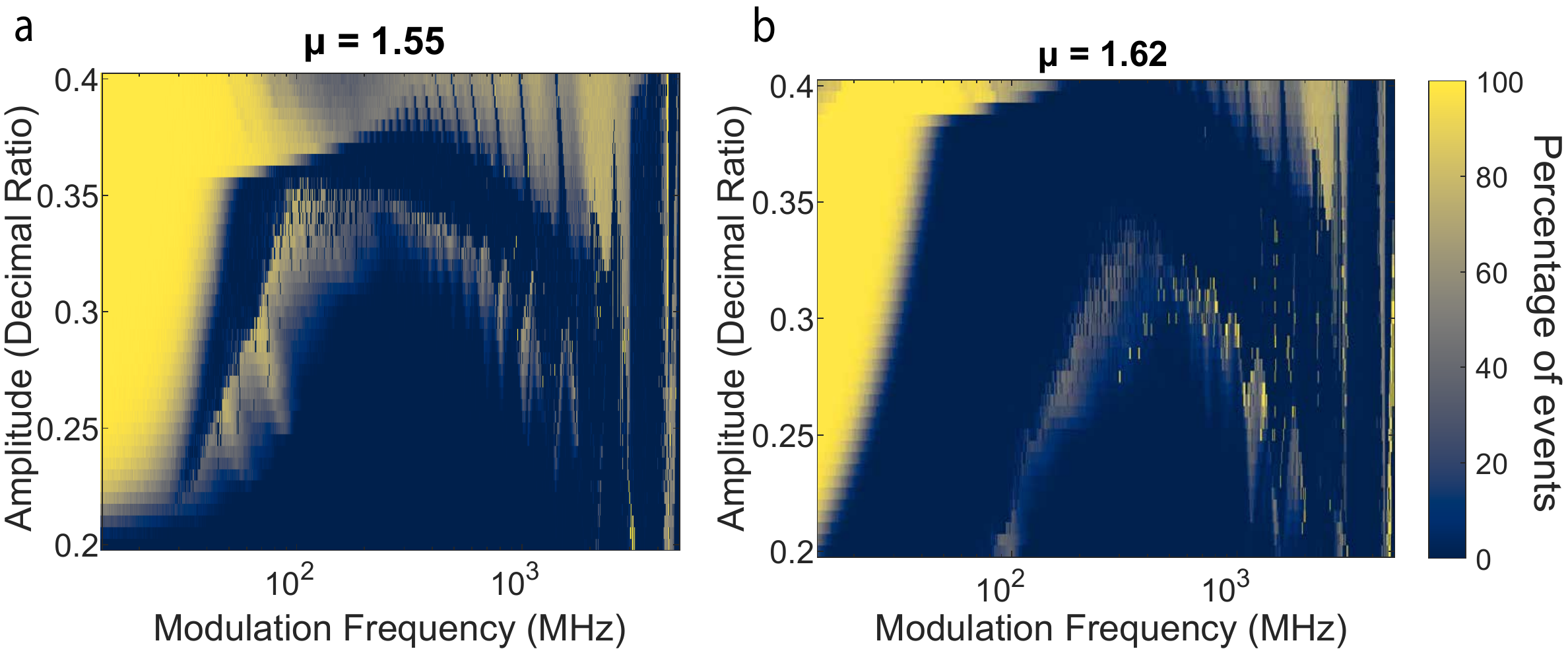}
  \caption{Polarization switching probability as a function of modulation frequency and amplitude for two bias currents: \(\mu=1.55\) (left) and \(\mu=1.62\) (right).}
  \label{fig:Amp_PS_colormap}
\end{figure}

The bifurcation diagram of Fig. \ref{fig:Bifurcation_diag} clearly displays a large hysteresis cycle accompanied by dynamical behavior when the current is increased. We can therefore naturally expect that the current offset and modulation amplitude will have a significant impact on the response of the laser, especially considering the complexity highlighted in the previous section.\\
We extend our numerical investigation to different bias currents around the switching point, as illustrated by the maps in Fig.~\ref{fig:Probability_colormap}.(a, b, c). To complement these results, we also provide the average power of both linear polarizations in each period in Fig.~\ref{fig:Probability_colormap}.(d,e). These maps unveil a complex picture of the polarization switching dynamics with many features. 
First, it is clear that periodic polarization switching can only be reliably achieved for low modulation frequencies, below 100 MHz, and within a specific current range between 1.45 and 1.7, as shown in Fig. \ref{fig:Probability_colormap}.(a). At higher frequencies, above 100 MHz and up to 2 GHz, and for lower currents between 1.35 and 1.45, switching doesn't occur systematically but only at a rate between 40 to 70 \%. For higher current values above 1.45, polarization switching happens only sporadically. 
Second, comparing Fig. \ref{fig:Probability_colormap}(b) and (c), it is interesting to see that the region where the X-LP polarization is dominant moves toward higher currents as the modulation frequency is increased. On the opposite, the Y-LP polarization is dominant at higher current only for low to moderate modulation frequencies between tens and about 400 MHz. For higher modulation frequencies, we observe a sharp line as the X-LP polarization looses its dominance. Although the underlying mechanism is not clear at this stage, such sharp changes could suggest a resonance of the modulation frequency with the internal response of the laser; notably, the frequencies involved are well below the conventional relaxation oscillation frequency, which suggests a link to polarization mode competition and spin-carrier dynamics rather than standard intensity modulation bandwidth limitations.\\
Last, from Fig. \ref{fig:Probability_colormap}.(d, e), we notice a clear asymmetry between X and Y polarizations: while X-LP is fully switched off when Y-LP is dominant, Y-LP is rarely fully suppressed. Indeed, the average power of Y-LP rarely goes below 0.1, which corresponds to about 10-20 \% of the total output power. This is consistent with the observations of Fig. \ref{fig:Time-Series}.(d) where dynamics remain visible on Y-LP even though X-LP is strongly dominant.\\
\begin{figure}
\centering
\includegraphics[width=\linewidth]{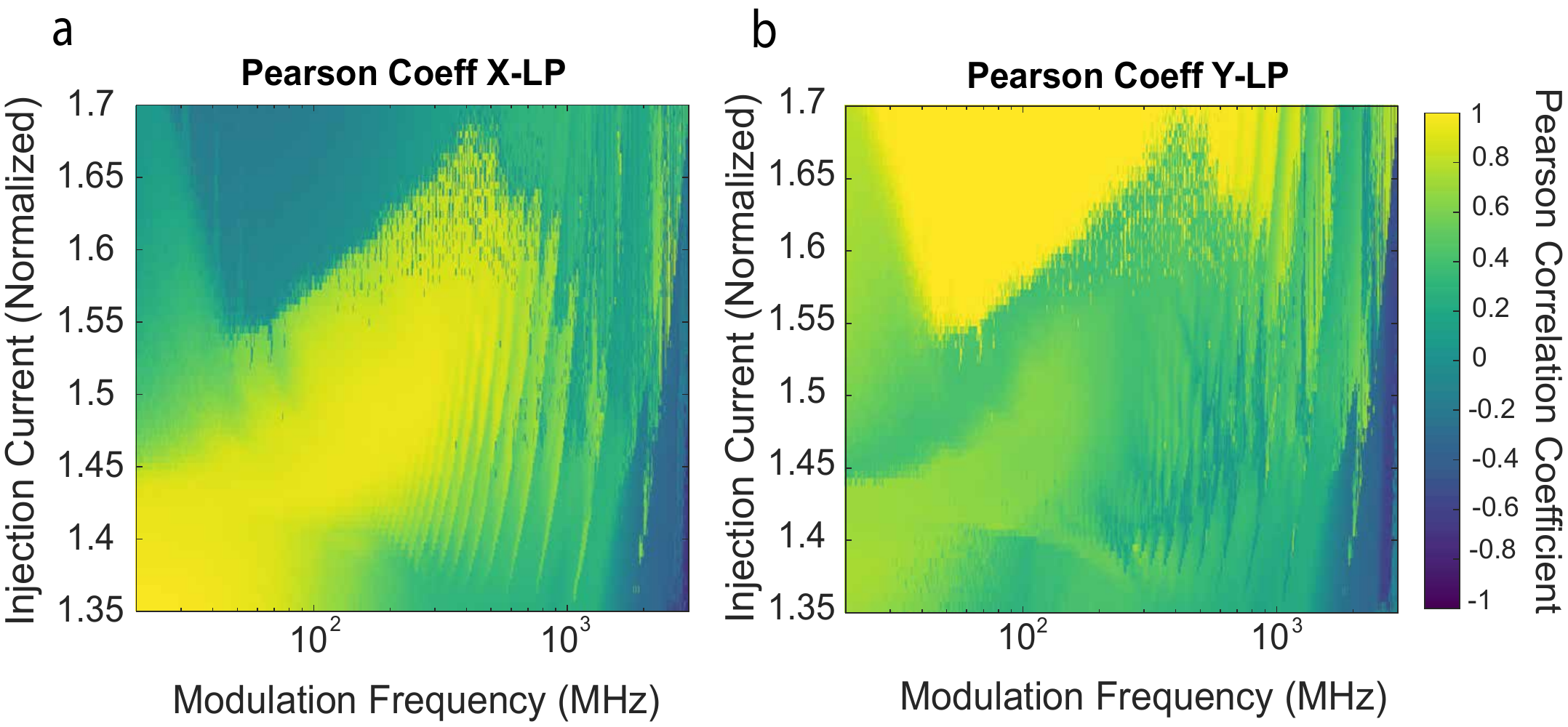}
\caption{Colormaps summarizing the evolution of the Pearson correlation coefficient between the current modulation waveform and the polarization-resolved time series, following the X linear polarization (panel a) and Y linear polarization (panel b), shown as a function of modulation frequency and injection current.}
\label{fig:PearsonCorr}
\end{figure}
In addition to the complexity of the dynamics, it is important to emphasize the lack of consistency in the laser response under current modulation. Indeed, in many configurations shown in Fig. \ref{fig:Probability_colormap}.(a), typically at higher modulation frequencies and lower currents, the laser exhibits irregular and non-systematic polarization transitions despite a purely sinusoidal driving signal. To quantify how faithfully the laser output follows the input modulation, we evaluate the Pearson correlation coefficient $P_{coeff}$ between the modulation waveform and the polarization resolved time series (Fig. \ref{fig:PearsonCorr}). The Pearson coefficient measures the linear similarity between two signals, with $P_{coeff}=1$ indicating perfect synchronous sinusoidal response, $P_{coeff}=0$ no linear correlation, and $P_{coeff}<0$ anti-correlated response. This metric is particularly suitable here because it directly quantifies waveform fidelity and phase consistency, unlike switching probability alone, which only captures event occurrence and not temporal synchronization.\\
From Fig. \ref{fig:PearsonCorr}.(b), high correlation ($P_{coeff}\approx1$) is observed only when the Y-LP state is dominant at low to moderate modulation frequencies, corresponding to regimes where the laser output closely follows the sinusoidal drive. In contrast, regions where $P_{coeff}<1$ indicate distorted, delayed, or irregular responses, reflecting incomplete synchronization between the modulation and polarization dynamics. In particular, in the stochastic switching region, the correlation remains low for both polarizations, confirming that the polarization evolution is only weakly linked to the driving signal and is instead dominated by intrinsic nonlinear dynamics and noise.
The evolution of the Pearson correlation coefficient colormaps (Fig. \ref{fig:PearsonCorr}) is consistent with the average power maps shown in Fig. \ref{fig:Probability_colormap}.(d,e), the comparison between the two indicates that regions of high correlation systematically coincide with clear dominance of one polarization mode. This confirms that strong synchronization with the modulation occurs only when a single mode remains dynamically stable.\\
Having established how modulation frequency and bias current influence synchronization and switching behavior, we next examine the role of the modulation amplitude in shaping the polarization dynamics (Fig. \ref{fig:Amp_PS_colormap}). To do so, we take two different injection current values at $\mu = 1.55$ and $1.62$ and plot the polarization switching probability as a function of the modulation frequency and amplitude. Immediately, we remark that for the lowest amplitudes, no polarization switching is observed, even at low frequencies. This is in line with the hysteresis cycle highlighted in the bifurcation diagram (Fig. \ref{fig:Bifurcation_diag}). Then, above a certain amplitude, polarization switching is systematically observed at low frequencies. Interestingly, this region expands towards higher modulation currents as the amplitude is increased. This is particularly significant above 0.35 and 0.37 for $\mu = 1.55$ and $1.62$ respectively. At higher frequencies, we observe the erratic patterns again, similar to those in previous maps. As expected, at a larger injection current $\mu = 1.62$, the laser is more stable as the region where Y-LP is dominant largely expands.\\
At larger amplitudes, an additional mechanism appears: the oscillating current periodically drops below the lasing threshold, resetting the device and triggering turn-on dynamics. In our data this occurs for \(A > 0.36\) at \(\mu = 1.55\) and for \(A > 0.38\) at \(\mu=1.62\). In that regime PS can be observed up to even higher modulation frequencies, but the temporal quality of the switching is degraded due to relaxation oscillations and non-ideal turn-on transients. Thus, while strong modulation extends the frequency reach of PS, it comes at the expense of clean cycle-to-cycle synchronization and period regularity. Such large modulation amplitude is similar to gain switching and the reported behavior appears to be consistent with the work of Quirce et al. \cite{Quirce:24_PolFluctuation_TurnOn}. It is interesting as it suggests that the stochastic switching observed at low amplitude could actually rely on a similar mechanism than the gain switching behavior reported in \cite{Quirce:24_PolFluctuation_TurnOn}.

\begin{figure}[tb]
  \centering
  \includegraphics[width=\linewidth]{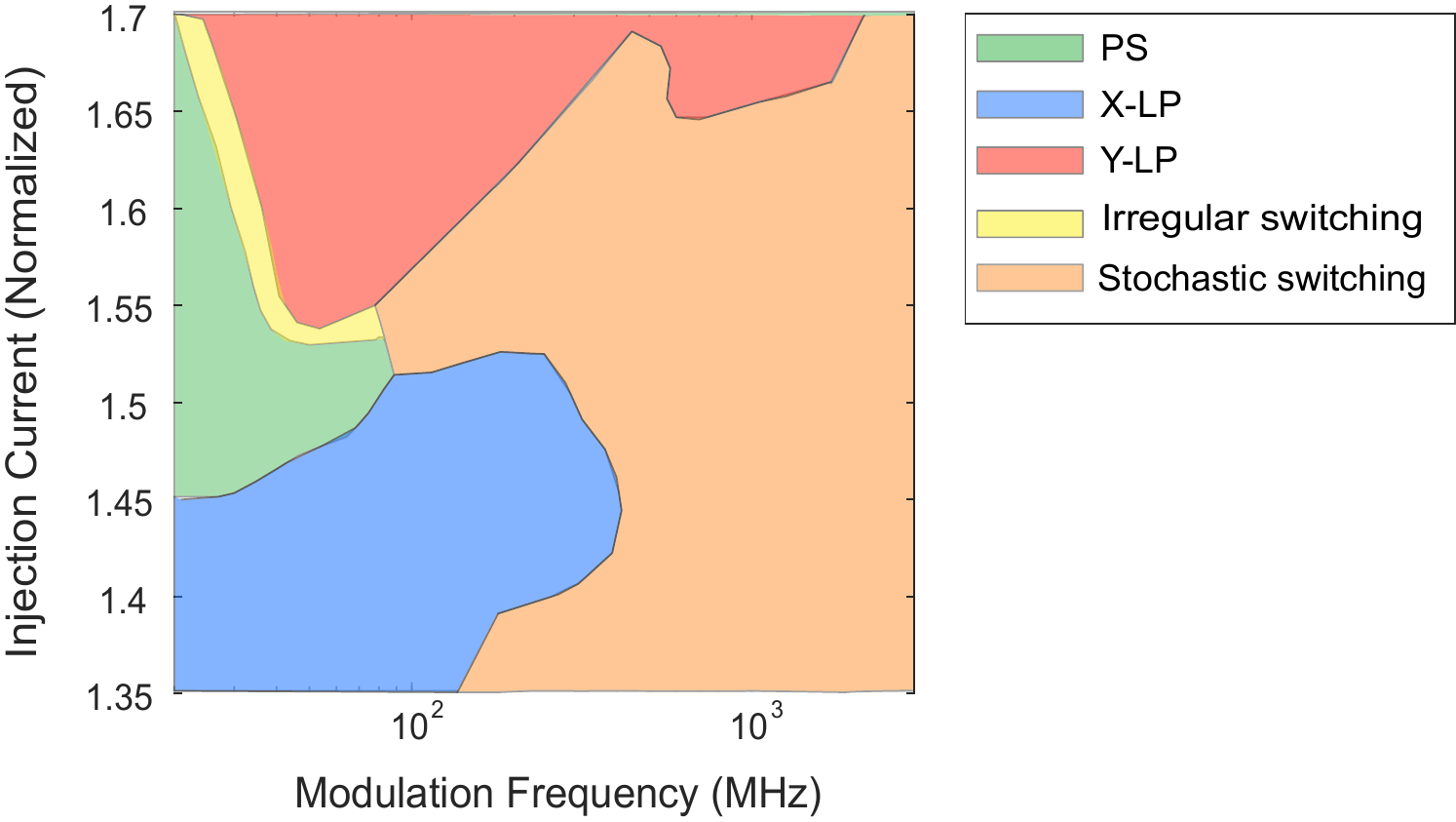}
  \caption{Summary regime map of the VCSEL polarization dynamics under sinusoidal current modulation near the polarization switching point. Colors indicate dominant dynamical behavior: deterministic polarization switching (PS), X-LP dominance, Y-LP dominance, and irregular or stochastic switching.}
  \label{fig:Regime_colormap}
\end{figure}

To consolidate the results discussed above, Fig. \ref{fig:Regime_colormap} provides a summarizing regime map of the VCSEL polarization dynamics under sinusoidal current modulation. The diagram condenses the multi-parameter analysis into a frequency–current plane, highlighting regions of deterministic polarization switching, single-polarization operation (with X-LP or Y-LP dominance), and irregular or stochastic switching behavior. This map offers a compact representation of the system response, capturing the transition from quasi-static switching to dynamically locked and stochastic regimes as the modulation frequency increases and the operating point is varied.

\section{Conclusion}
In this work, we have investigated the polarization dynamics of Vertical-Cavity Surface-Emitting Lasers under sinusoidal signal modulation near the polarization switching point. Using the Spin-Flip Model, we performed detailed numerical simulations to characterize the laser polarization response across a range of modulation frequencies.\\ 
Our results reveal distinct dynamic regimes depending on the proximity to the PS point, as well as the modulation frequency and amplitude. At low frequencies, polarization switching follows the input signal quasi-statically. Hysteresis appears near the polarization switching threshold. As the frequency increases, the system transitions to dynamic polarization oscillations, including deterministic switching, partial switching, polarization locking, and stochastic polarization oscillations, reflecting the onset of nonlinear dynamics. We classified the response into several characteristic regimes, providing a framework for understanding how sinusoidal modulation can be used to control the polarization state in VCSELs. These insights can be relevant for applications in polarization-encoded communications and advanced photonic circuits where polarization stability and switching control are crucial. Beyond their fundamental interest, the identified regimes of deterministic switching, polarization locking, and stochastic behavior under sinusoidal current modulation provide practical guidance for polarization-sensitive systems operating at frequencies up to hundreds of $\text{MHz}$. Our regime maps highlight frequency–current windows where we can systematically control the polarization state through the current modulation, regions where dynamics decouple from the modulation and become stochastic, and regions that can potentially be harnessed as a source of polarization-defined high-speed random number generation.\\
Together, modulation amplitude and bias current provide complementary control: amplitude shifts the regime thresholds in frequency, while bias current adjusts the width of stable operational windows. This tunability could offer practical means to tailor the laser’s polarization response for applications requiring precise control over switching dynamics, mode selection, and synchronization with external signals. Increasing the amplitude effectively enhances the periodic forcing, allowing the system to overcome the polarization switching barrier more easily. Consequently, the threshold frequencies separating the dynamic regimes shift: higher modulation amplitudes push the transition to partial switching or polarization locking toward higher frequencies, thereby extending the accessible range of cycle-synchronized deterministic switching. Conversely, smaller amplitudes cause transitions to occur at lower frequencies, compressing the operational window for deterministic switching.\\
Finally, we remark an interesting connection with the reported polarization dynamics of gain-switched VCSELs in \cite{Quirce:24_PolFluctuation_TurnOn}, which shows the random selection of a polarization mode at turn on. Although further work is required to fully confirm this aspect, it seems that modulation of the current closer to the switching point could lead to a similar random outcome of the polarization, which may be exploited in the same fashion for random number generation. 

% \section*{Acknowledgment}
% The authors acknowledge the support of the Research Foundation - Flanders (FWO, Grant number G020621N ), the European Research Council (ERC, Starting Grant COLOR’UP 948129, MV) and the METHUSALEM program of the Flemish government.

% \section*{Data availability statement}
% All data and materials necessary to reproduce the results are available in Zenodo repository at https://doi.org/10.5281/zenodo.18456280.

\bibliographystyle{IEEEtran}
\bibliography{Biblio.bib}
%\input{output.bbl}

% \printbibliography

\end{document}